\documentclass[lettersize,journal]{IEEEtran}
\usepackage{amsmath,amsfonts}
\usepackage{algorithmic}
\usepackage{algorithm}
\usepackage{array}
\usepackage[caption=false,font=normalsize,labelfont=sf,textfont=sf]{subfig}
\usepackage{textcomp}
\usepackage{stfloats}
\usepackage{url}
\usepackage{verbatim}
\usepackage{graphicx}
\usepackage{cite}
\usepackage{xcolor}
\begin{document}

\title{An Analytical Model for SNR Prediction in  Coherent Systems after generic Jones Matrixes}

\author{Giuseppe~Rizzelli, Pablo~Torres~Ferrera, and~Roberto~Gaudino,~\IEEEmembership{Senior Member,~IEEE}
\thanks{G. Rizzelli is both with LINKS Foundation, Torino, Italy and the Department
of Electronics and Telecommunications, Politecnico di Torino, Torino, e-mail: giuseppe.rizzelli@linksfoundation.com.}
\thanks{P. Torres and R. Gaudino are with the Department
of Electronics and Telecommunications, Politecnico di Torino, Torino,
Italy.}}



\maketitle

\begin{abstract}
In this letter, we propose the extension of a previously presented analytical model for the estimation of the signal-to-noise ratio (SNR) at the output of an adaptive equalizer in coherent optical transmission systems when transmission is modeled as a generic 2x2 matrix transfer functions, to be applied to polarization multiplexed communications based on advanced modulation formats (PM-QAM). We present the model and then we extensively test its accuracy on an possible application environment.  Our findings show a remarkable agreement, with maximum SNR discrepancies of about 0.5 $dB$ between time-domain simulations and analytical results.
\end{abstract}


\section{Introduction}
\label{introduction}
\IEEEPARstart{C}{oherent} detection (CD) coupled with digital signal processing (DSP) is the ultimate choice in modern high-speed optical communications systems \cite{IMDDvsCOH,Coherent}. Coherent systems rely on advanced modulation formats to achieve very high bitrates by taking advantage of three main coherent signal features: intensity, phase and polarization. At the receiver side, sophisticated DSP algorithm are required to compensate for the (linear and nonlinear) impairments introduced by the optical transmisison channel. Typical functions of the DSP at the receiver include: clock recovery, carrier phase recovery, polarization recovery and adaptive feed-foorward equalization equalization \cite{Savory}.
In several situation, such as for physical-layer aware optical network dimensioning  \cite{curri} using software planning tools, ultra-fast estimation of physical layer performance is needed, typically based on available optical SNR (OSNR) at the input of the receiver and on the subsequent   SNR obtained at the output of the equalizer for a given received optical power level, which is then directly related to the system BER. While in coherent system the relation between OSNR and SNR is trivial for frequency flat channels, it becomes analytically more complex when there is a frequency dependence on the optoelectronics (for instance due to optical frequency-selective filters/switches and/or electrical bandwidth limitations in the transceivers) and, even more, when polarization dependent effects are present.
In simulation, the output SNR Tcan be estimated with good accuracy through time-domain simulations that consider also the details of the DSP \cite{Savory}, but in some applications this is too CPU-time consuming. This situation is for instance typical either for real-time network optimization or for simulation scenarios requiring multi-dimensional parameters swept such as, for instance, in a Monte Carlo approach. 

In this paper, we thus present an analytical model that allows to obtain SNR estimation at the output of the coherent receiver DSP assuming: 
\begin{itemize}
    \item PM-QAM transmission
    \item a linear optical channel, with a generic 2x2 matrix transfer functions
    \item additive Gaussian noise at the receiver (not necessarily flat)
    \item a coherent receiver implementing today typical DSP algorithms
\end{itemize}
We note that in several performance assessments in long-haul network dimensioning, such as for \cite{curri}, this is the common situation since, thanks to the well-known Gaussian-Noise models \cite{Poggiolini_GN}, fiber nonlinearities are actually modelled through additive Gaussian noise at the receiver, while signal propagation is linearly modelled.

For a single polarization QAM transmission, it is possible to use the analytical model described is \cite{Fischer} which assumes an infinitely long linear minimum mean-squared error (MMSE) equalizer and computes the achievable SNR at the equalizer output as a function of the spectral signal-to-noise ratio (or channel SNR function \cite{HU98}).

In this manuscript we thus extend this model \cite{Fischer} to make it applicable to PM-QAM, i.e. to a generic polarization multiplexed transmission and a generic channel including both frequency and polarization dependence. 

 The reminder of this manuscript is organized as follows: in Section \ref{model} we describe the analytical model while in Section \ref{results} we address its accuracy by presenting an application scenario where we compare performance estimation obtained with  the analytical model and with a detailed time-domain simulation on a PM-16QAM signal. 
 In Section \ref{conclusion} we discuss the implications of the presented results and draw some conclusions.

\section{The Analytical Model}
\label{model}
We start by observing that in coherent systems, the relevant noise source is typically additive Gaussian noise on the $\hat{x}$ and $\hat{y}$ optical fields. This is well know for long-haul optically amplified systems (due to optical amplifier ASE noise) but also  to short-reach links  without optical amplification since, as we show for instance in \cite{scaling_laws}, also in this case the relevant noise (typically a combination of shot noise and trans-impedance amplifier thermal noise) is to a good approximation again Gaussian and additive on the four electrical signals proportional to the $\hat{x}$ and $\hat{y}$ optical fields.
  The system we want to study can thus be numerically simulated as the cascade of the following elements:
\begin{enumerate}
    \item a PM-QAM transmitter generating the signal $\vec{E}_{in}(t)$ in the transmitter SMF fiber;
    \item an optical propagation links that can be modelled as a totally generic [2x2] frequency dependent transfer function $\bar{\bar{H}}(f)$ acting on the transmitted PM-QAM signal;
    \item a coherent receiver acting on $\vec{E}_{out}(t)=\boldsymbol{H}(f)\vec{E}_{in}(t) +\vec{n}(t)$ where $\vec{n}(t)$ is the additive Gaussian noise (possibly non-flat).
\end{enumerate}

To predict the system performance, starting from a certain end-to-end transfer function, we can thus use the analytical model presented in \cite{Fischer}, derived under the assumption of additive white Gaussian noise (AWGN) channel with a generic transfer function $H_c(f)$. In \cite{Fischer} the SNR at the output of an infinitely long adaptive MMSE equalizer is computed from the spectral SNR at the receiver input, in the single polarization case as:
\begin{equation}\label{eq:SNR}
    SNR = \frac{1}{T \cdot \int_{-\frac{1}{2T}}^{\frac{1}{2T}} \frac{1}{\overline{SNR}(f)+1} df}
\end{equation}
where $T$ is the symbol time (i.e. the inverse of the baud rate) and $\overline{SNR}(f)$ is the folded version of the spectral SNR, defined as:
\begin{equation} \label{eq:SNR_tilde}
    \overline{SNR}(f) =  \sum_{\mu} SNR(f - \frac{\mu}{T})
\end{equation}
The spectral SNR is:
\begin{equation}\label{eq:SNR_f}
    SNR(f) = \frac{T \sigma_a^2 |H_T(f)H_C(f)|^2}{N_0(f)}
\end{equation}
where $\sigma_a^2$ is the transmitted signal power, $H_T(f)$ is the transfer function of the transmitter shaping filter (meaning that the transmitted useful signal power spectral density is proportional to $\sigma_a^2 \cdot |H_T(f)|^2 $ ), a squared root raised cosine filter with $0.2$ roll-off factor in our study
$N_0(f)$ is the equivalent noise power spectral density at the input of the receiver.

Since our goal is to apply this approach to the double polarization case and study polarization multiplexed coherent systems, we extend this model to include generic [2x2] transfer function matrix $\boldsymbol{H}(f)$ thus being able to describe any linear optical channel (including arbitrary frequency and polarization dependence). The transmission system is thus described as:
\begin{equation}\label{eq:system}
\begin{bmatrix}
E^{x}(f)\\
E^{y}(f)
\end{bmatrix}_{RX} = \boldsymbol{H}(f) \cdot \begin{bmatrix}
E^{x}(f)\\
E^{y}(f)
\end{bmatrix}_{TX}
\end{equation}
where
\begin{equation}\label{eq:system2}
\boldsymbol{H}(f) = \begin{bmatrix}
H^{xx}(f) & H^{xy}(f)\\
H^{yx}(f) & H^{yy}(f)
\end{bmatrix}
\end{equation}
\begin{figure}[tbp]
\centering
\includegraphics[width=1\linewidth]{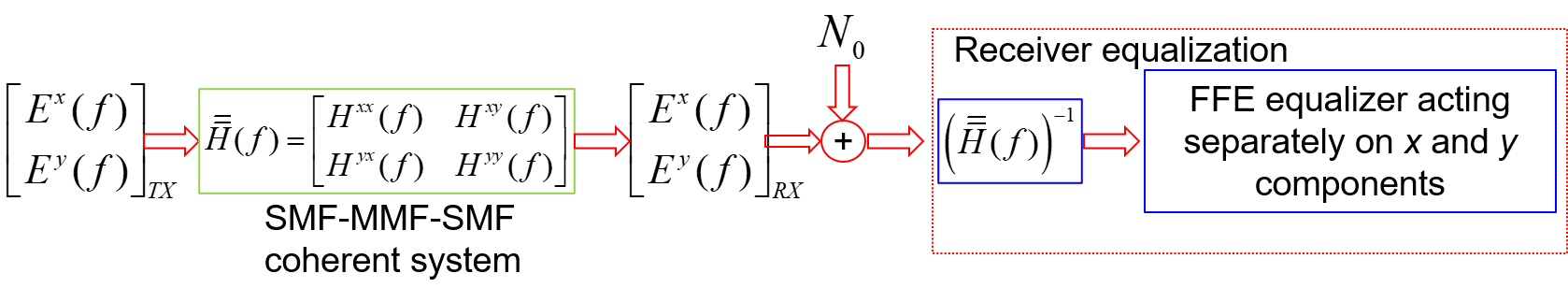}
\caption{Receiver structure.}
\label{fig:receiver}
\end{figure}

The receiver schematic is shown in Fig. \ref{fig:receiver}. In our derivation, we assume that the $\boldsymbol{H}(f)$ matrix is invertible inside the useful signal bandwidth and that, only for the target of analytical performance estimation, the receiver equalizer thus has the following structure:
\begin{itemize}
    \item A first 2x2 stage $\boldsymbol{H}(f)^{-1}$ that compensates the polarization rotation introduced by the channel Jones matrix at that frequency. At the output of this block, the two transmitted PM signals are thus properly re-aligned on the proper $\hat{x}$ and $\hat{y}$, without mutual crosstalk .
    \item after polarization crosstalk has been removed, standard FFE equalization is applied on each of the two PM components separately.
\end{itemize}

At the input of the feed forward equalizer (FFE) we thus have that the transmitted signal has been perfectly retrieved (zero crosstalk on the useful signal), but the Gaussian noise components have been enhanced by the cascade of the two transfer functions ($\boldsymbol{H}(f)^{-1}$ and the one introduced by the FFE equalizer):
\begin{equation}\label{eq:ffe}
\begin{bmatrix}
E^{x}(f)\\
E^{y}(f)
\end{bmatrix} = \begin{bmatrix}
E^{x}(f)\\
E^{y}(f)
\end{bmatrix}_{TX} + \boldsymbol{K}(f) \cdot \begin{bmatrix}
n^{x}(f)\\
n^{y}(f)
\end{bmatrix}_{RX}
\end{equation}
where:
\begin{equation}\label{eq:K}
\boldsymbol{K}(f) = \boldsymbol{H}(f)^{-1} = \begin{bmatrix}
K^{xx}(f) & K^{xy}(f)\\
K^{yx}(f) & K^{yy}(f)
\end{bmatrix}
\end{equation}

Translating Eq. \eqref{eq:ffe} in terms of power spectral densities we get:
\begin{equation}\label{eq:PSD}
\begin{bmatrix}
P^{x}(f)\\
P^{y}(f)
\end{bmatrix} = \begin{bmatrix}
P^{x}(f)\\
P^{y}(f)
\end{bmatrix}_{TX} + \begin{bmatrix}
|K^{xx}(f)|^2 + |K^{xy}(f)|^2\\
|K^{yx}(f)|^2 + |K^{yy}(f)|^2
\end{bmatrix} \cdot N_0
\end{equation}
so that finally:
\begin{equation}\label{eq:SNR_final}
\begin{array}{c}
SNR_x(f) = \frac{P^x(f)}{(|K^{xx}(f)|^2 + |K^{xy}(f)|^2) \cdot N_0} \\ \\
SNR_y(f) = \frac{P^y(f)}{(|K^{yx}(f)|^2 + |K^{yy}(f)|^2) \cdot N_0}
 \end{array}
\end{equation}
This is the main result of our paper, since it allows to analytically derive the spectral $SNR(f)$ on each of the two $\hat{x}$ and $\hat{y}$ useful signal components, which can finally be inserted in Eq. \eqref{eq:SNR} to estimate the $SNR_x$ and $SNR_y$  performances on each of the two QAM signals and, consequently, the two resulting BERs.

\section{Application example and model accuracy check}
\label{results}

In \cite{OFC2022} we have presented an experimental analysis of an SMF-MMF-SMF coherent system, that is a possible configuration for next generation CD-based intra-datacenter (IDC) links, where the existing multi mode fiber (MMF) infrastructure of a data center can be reused in combination with coherent technology to overcome the speed limitations imposed by traditional intensity modulation and direct detection- (IMDD-) based solutions. In this context we assume that the coherent transceiver developed for this application will still be coupled to a short piece of single mode fiber (SMF) fiber both at the output of the transmitter and at the input of the receiver. In this configuration, only the fundamental LP\textsubscript{01} mode can propagate along the SMF sections, whereas high order modes can be excited inside the MMF. Thus, the resulting end-to-end transfer function of the system becomes strongly frequency dependent due to the relative delays of the MMF modes and to the birefringence seen by each mode in the MMF. It can be shown that the relation between the $\vec{E}_{in}$  field of the LP\textsubscript{01}  of the first SMF fiber (i.e. the signal generated by the coherent TX) and the $\vec{E}_{out}$ field  at the output of the last SMF section (i.e. the signal received by the coherent RX) is described by:

\begin{equation}\label{eq:smf_mmf_smf}
\vec{E}_{out}(t) = \sum_{j=0}^{M-1}\rho_j^{in}\boldsymbol{J}_j \cdot \vec{E}_{in}(t-\tau_j)\rho_j^{out}
\end{equation}

where $M$ is the total number of modes of the MMF, $j$ is the index of the $j$th mode of the MMF, $\rho_j^{in}$ is the coupling coefficient between the LP\textsubscript{01} mode of the input SMF and the $j$th mode of the MMF, $\rho_j^{out}$ is the coupling coefficient between the $j$th mode of the MMF and the LP\textsubscript{01} of the output SMF, $\boldsymbol{J}$ is the unitary random Jones matrix  that takes into account "per mode" fiber birefringence and $\tau_j$ is the modal delay of the $j$th mode inside the MMF. Without loss of generality, and only to simplify the equations, we referred all delays and birefringence to those of the fundamental LP\textsubscript{01} mode, therefore $\tau_0=0$. Moreover, the $\rho_j$ parameters can be calculated using the analytical model presented in \cite{rho_model}.

By Fourier transforming Eq.~\eqref{eq:smf_mmf_smf}, we conclude that the transmitted signal undergoes the following [2x2] frequency-dependent transfer function matrix:
\begin{equation}\label{eq:matrix_transfer_function}
\boldsymbol{H}(f)= \sum_{j=0}^{M-1}\rho_j^{in}\boldsymbol{J}_{j} e^{-j 2 \pi f \tau_j} \rho_j^{out}
\end{equation}

In this Section we apply the analytical model presented in Section \ref{model} to estimate the SNR of polarization multiplexed coherent systems whose transfer function is affected by a strong frequency dependence due to the modal delay and birefringence inside a MMF. To do so, we generate a variety of frequency responses by randomly generating Jones matrices for each mode of the MMF, and we compare the results obtained analytically to the output of a time domain simulator based on bit error counting. In this case we assume $H_T(f)$ is a squared root raised cosine filter with $0.2$ roll-off factor.

\begin{figure}[htbp]
\centering
\includegraphics[width=1\linewidth]{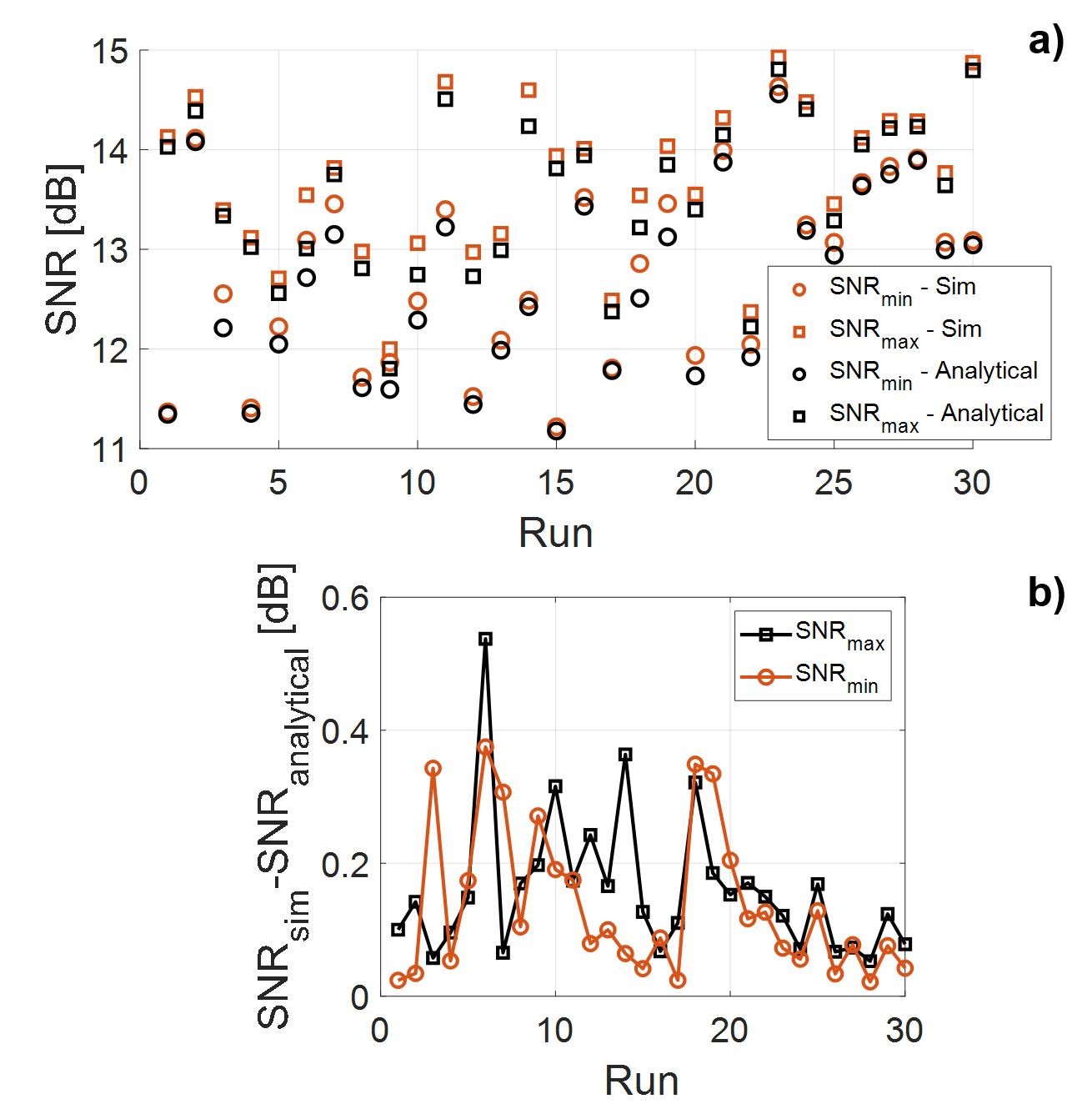}
\caption{a) Minimum (circles) and maximum (squares) SNR for a 25 GBaud PM-16QAM SMF-MMF-SMF configuration with 100 m MMF fiber and no MMF-MMF connectors, obtained through time-domain simulations (red) and the proposed analytical model (black). b) Difference between SNR obtained through time-domain simulations and the proposed analytical model.}
\label{fig:model_check}
\end{figure}

As a check on the validity of the proposed performance estimation method, Fig. \ref{fig:model_check} shows the results obtained through the analytical model compared to the time-domain simulations, for a 25 $GBaud$ PM-16QAM SMF-MMF-SMF system with 100 $m$ MMF fiber and 30 randomly generated Jones matrices accounting for variable birefringence. In Fig. \ref{fig:model_check}a both the minimum and maximum SNR values are shown, highlighting very small discrepancies between the two methods. Modal dispersion changes the way all the modes inside the MMF couple to the $LP_{01}$ of the output SMF. This sort of interferometric effect is scrambled by the randomly generated birefringence and causes the SNR to fluctuate. Over the considered 30 runs, the SNR varies by more than 3.5 $dB$. The difference between the SNR computed with the two methods is calculated in $dB$ and shown in Fig. \ref{fig:model_check}b for both variables. For most of the 30 runs, the analytical model yields less than 0.2 $dB$ error with respect to the simulations in the time domain. Moreover, the estimation is always conservative, with the model producing always slightly worse results, as expected from a non-perfect zero-forcing-like kind of equalizer, due to noise enhancement. 

\section{Conclusions}
\label{conclusion}

We have obtained an analytical model to assess the performance of optical coherent systems for generic polarization and frequency dependence on the transmission channel and generic additive noise, assessing its accuracy compared to much more CPU-time consuming time domain simulations in a specific example scenario. We point out that our model can anyway be applied  in several other scenarios, as we mentioned in the Introduction. In fact, many modern optical core network planning tools \cite{curri} are today
based on equivalent linear models where it can be useful to consider polarization and frequency dependence. Just as an example, this can be true in analysis considering generic (possibly polarization dependent) Reconfigurable Optical Add-Drop Multiplexers (ROADMs)  \cite{ROADM_application} with realistic (i.e. non ideal) transfer functions.

\end{document}